\documentclass[prb,aps,amsmath,preprint,showpacs]{revtex4}

\usepackage{graphicx}

\usepackage{dcolumn}

\usepackage{bm}

\begin{document}

\title{High pressure structural study  
of fluoro perovskite CsCdF$_3$  upto 60 GPa: A combined experimental and theoretical study} 
 
\author{G. Vaitheeswaran$^{1*, 2}$, V. Kanchana$^{1}$, Ravhi. S. Kumar$^{3}$, \\ 
A. L. Cornelius$^{3}$, M. F. Nicol$^{3,4}$, A. Svane$^{5}$, N. E.
Christensen$^{5}$ and O. Eriksson$^{6}$} 

\affiliation{$^{1}$Applied Materials Physics, Department of Materials Science and Engineering, 
Royal Institute of Technology, Brinellv\"agen 23, 100 44 Stockholm, Sweden \\
$^{2}$Advanced Centre of Research in High Energy Materials (ACRHEM), 
University of Hyderabad, Hyderabad 500 046, Andhra Pradesh, India \\
$^{3}$High Pressure Science and Engineering Center and Department of Physics, University of Nevada, 
Las Vegas, Nevada 89154, USA \\
$^{4}$Department of Chemistry, University of Nevada,Las Vegas, Nevada 89154, USA\\
$^{5}$Department of Physics and Astronomy, Aarhus University, DK-8000 Aarhus C, Denmark\\
$^{6}$Department of Physics and Materials Science, Uppsala University, Box 590, SE-751 21, Uppsala, Sweden}

\date{\today}

\begin{abstract}
The structural behaviour of CsCdF$_3$ under pressure  
is investigated by means of theory and experiment. 
High-pressure powder x-ray diffraction experiments were
performed up to a maximum pressure of 60 GPa using synchrotron 
radiation. 
The cubic $Pm\bar{3}m$ crystal symmetry persists throughout this pressure range.
Theoretical calculations were carried out using the full-potential linear muffin-tin orbital method
within the local density approximation and
the generalized gradient
approximation for exchange and correlation effects. 
The calculated ground state properties  -- the equilibrium lattice constant, bulk modulus 
and elastic constants -- are in good agreement with experimental results. 
Under ambient conditions, CsCdF$_3$ is an indirect gap insulator 
with the gap increasing under pressure.
\end{abstract}

\pacs{62.50.-p, 61.05.cp, 62.20.de, 71.15.Nc}
\maketitle

\section {Introduction} 
Ternary fluorides with the perovskite crystal structure 
have been extensively studied over several 
decades, as they have several    
potential applications because of their
optical properties\cite{Paus,Hua},
high-temperature super-ionic behaviour\cite{Chadwick}, and 
physical properties, such as  
ferroelectricity\cite{Hull}, antiferromagnetism\cite{Julliard}, and 
semiconductivity\cite{Daniels}. 
The application of 
CsCdF$_3$ in the field of luminesence\cite{Villacampa,Villacampa2,Barriuso} has motivated 
 several experimental and theoretical investigations of defect structures involving 3d transition 
metal ions.\cite{Owen,Adam,Minner,Takeuchi,Springis,Yao,Hong,Wu,Aramburu}.
Elastic properties, including thermal expansion  and acoustical  measurements of higher order
elastic constants, have been reported\cite{Rousseau,Damodar,Philip,Fischer}. 
Perovskite fluorides may exhibit structural phase transitions as function of temperature and
pressure.\cite{Ross,Zhao,Zhao2}
The present work is a combined theoretical and experimental study of 
the ground state and high-pressure properties of CsCdF$_3$. 
We present the equation of state resulting from high-pressure 
diamond-anvil cell experiments up to 60 GPa.
The ideal cubic structure is preserved over this entire pressure range.
We also present 
the equation of state, the elastic constants 
and the electronic structure as obtained
from theoretical calculations based on density functional theory
using two different approximations for the exchange-correlation functional. 

The remainder of the paper is organized 
as follows. Details of the computational 
method as well as details of the experimental setup are 
outlined in section II. The measured and calculated equations of state are presented
in section III together with calculated ground state properties
and elastic properties.
The electronic structure 
and the pressure variation of the band gap are discussed in section IV.  
Finally, conclusions are given in section V.

\section {computational and experimental details}
\subsection{The electronic structure method}
The total energies and basic ground state properties of CsCdF$_3$ were 
calculated by the
all-electron linear muffin-tin orbital (LMTO) method \cite{OKA75} 
in the full-potential implementation of Ref. \onlinecite{Savrasov}. 
In this method, the crystal volume is split into two
regions: non-overlapping muffin-tin
spheres surrounding each atom and the
interstitial region between these spheres. We used a double $\kappa$
spdf LMTO basis to describe the valence bands, {\it i. e.}
atom-centered Hankel functions with characteristic decay rate denoted by $\kappa$ are matched to 
a linear combination of products of numerical radial function and spherical harmonics within the muffin-tin spheres.
The calculations included the 5s, 5p, 6s,  and 5d partial waves for cesium, the
5s, 5p, and 4d partial waves for cadmium, and the 2s and 2p partial waves for fluorine. 
The exchange correlation potential was 
calculated within the local density approximation (LDA)\cite{Vosko} as well as 
the generalized gradient approximation (GGA) scheme\cite{Perdew}.
The charge density and potential inside the muffin-tin spheres 
were expanded in terms of spherical
harmonics up to $l_{max}$=6, while in the
interstitial region, they were expanded in plane
waves, with 14146 waves (energy up to 156.30 Ry) 
included in the calculation. Total energies
were calculated as a function of volume for a 
(16 16 16) k-mesh containing 165 
k-points in the irreducible wedge of the Brillouin zone and 
were fitted to a second order
Birch-Murnaghan equation of state\cite{Birch} to obtain the ground state properties.

The elastic constants were obtained from the variation of the total energy under 
volume-conserving strains, as outlined in Refs.  \onlinecite{Oxides}
and \onlinecite{srcl2}.

\subsection{Experimental details}
Polycrystalline samples of CsCdF$_3$ were synthesized by the solid state reaction method as described elsewhere.\cite{Chadwick, zhao,smith} 
For high pressure powder diffraction experiments, samples with a few ruby
chips were 
loaded in a Mao-Bell type diamond anvil cell in a rhenium gasket (135 $\mu$m  hole diameter and pre-indentation thickness 65 $\mu$m). Pressure 
was generated with 325 $\mu$m  culet diamonds. Silicone fluid was
 the pressure transmitting medium.\cite{Shen} X-ray diffraction experiments were 
performed at Sector 16 ID-B, HPCAT, at the Advanced Photon Source. The 
pressure inside the diamond anvil cell was determined by the standard
ruby luminescence method.\cite{Mao} 
The incident monochromatic x-ray wavelength 
was $\lambda$ = 0.36798 \AA.  The x-ray diffraction patterns were recorded on an imaging plate with a typical 
exposure time of 10-20 s, with an 
incident beam size of 20 x 20 $\mu$m. The distance between the 
sample and the detector was calibrated using a CeO$_2$ standard.
The patterns were integrated using the Fit2D software program\cite{Hammersley} 
and the cell parameters were obtained using the JADE package.

The evolution of diffraction patterns as a function of pressure is shown in Fig. 1. At ambient conditions the cell 
parameter a = 4.4669(7) \AA\ obtained experimentally agrees well
with earlier reports provided 
in the Inorganic Crystal Structure Data base (ICSD).\cite{Hoppe}
Thermal expansion and temperature dependent structural experiments on 
CsCdF$_3$ were
performed earlier by Reddy et al.\cite{Damodar} They found that the
coefficient of thermal expansion 
increases with 
temperature. Rousseau et al.\cite{Rousseau} investigated the
crystal structure and elastic constants of RbCdF$_3$, TlCdF$_3$ and CsCdF$_3$ 
using x-ray diffraction and Raman techniques at varying  temperatures. They
found tetragonal distortions in RbCdF$_3$ and TlCdF$_3$ at 124 K and 191 K
respectively. No structural phase transition were reported in CsCdF$_3$ in 
their temperature dependent study. Investigations by Haussuhl\cite{Haussuhl} showed that
temperature induced transitions can be observed in the pseudocubic
perovskite CsCd(NO$_2$)$_3$, which undergoes a phase transition from $R3$
to $Pm3m$ symmetry around 464 K.\cite{Haussuhl} These reports indicate that the
cubic structure of CsCdF$_3$ is relatively stable in comparison with
RbCdF$_3$, TlCdF$_3$ and KMgF$_3$ for a wide range of temperature from
 4 K to 300 K.

Fischer\cite{Fischer} investigated
 the pressure dependence of the second order elastic constants of CsCdF$_3$ and 
 related fluoroperovskites. He reported a decrease in the force constant in RbCdF$_3$, 
 which is connected with the instability of the F ions. Such
 instability has not been noticed for CsCdF$_3$, however a possible phase transition at high pressure was suggested.\cite{Fischer} So far, no high pressure experiments
 have been reported to ellucidate the structural behaviour of CsCdF$_3$.
In a previous work we showed that cubic KMgF$_3$ is stable 
up to 40 GPa\cite{kmgf3}. If any structural phase transition is 
to be induced in CsCdF$_3$, one 
would expect
that   pressures above 40 GPa are required. However, the diffraction patterns collected in our experiments show no pressure induced structural 
transformations up to 60 GPa. That is, similarly to KMgF$_3$, 
 the cubic phase of CsCdF$_3$ remains stable up to the highest pressures achieved in our experiments.

The bulk modulus was obtained by fitting the pressure-volume data to a second order Birch-Murnaghan equation of state:\cite{Birch}
\begin{equation}
 P = \frac{3}{2} B_0\left[(V/V_0)^{-7/3} - (V/V_0)^{-5/3}\right] \cdot \{1+3/4(B_0'-4)[(V/V_0)^{-2/3} - 1]\},
\end{equation}
 where $B_0$ is the bulk modulus and $B_0'$ its pressure derivative. Least square fitting resulted in a $B_0$ of 79(3) GPa, with a $B_0'$ of 3.8. 
Thus, CsCdF$_3$ has a slightly higher bulk modulus than  KMgF$_3$ ( $B_0=71.2$ GPa, $B_0'=4.7$)\cite{kmgf3}

\begin{figure}
\label{figXRD}
\begin{center}
\includegraphics[height=100mm,width=100mm,angle=0]{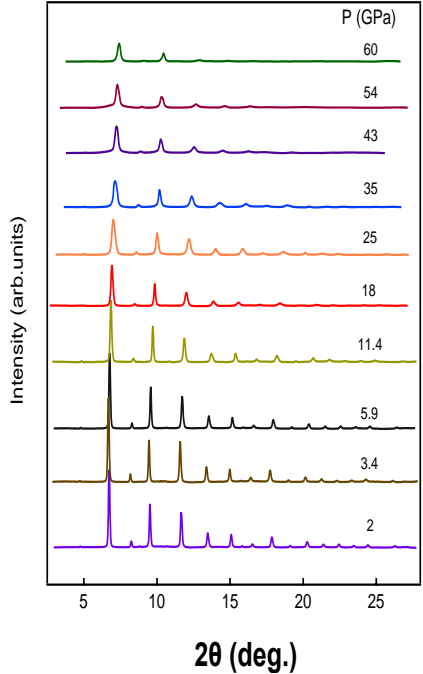}\\
\caption{ (Color online) Powder x-ray diffraction patterns of CsCdF$_3$ recorded at 
ten  pressures up to
 60 GPa}
\end{center}
\end{figure}

\section{Ground state and Elastic properties}
Powder x-ray diffraction patterns collected at several pressures are shown in Figure 1.
On compression, the diffraction patterns remain unchanged up to 60 GPa, 
except for the shifts of diffraction lines caused by the decreasing lattice
constant and the broadening due to the intensity decrease with increasing
pressure.
This implies that no structural transformations occur up to 60 GPa in CsCdF$_3$.
Figure 2 shows the measured equation of state of 
CsCdF$_3$ and compares it with theoretical curves calculated within the 
LDA and GGA. 
As is typical, the LDA leads to slightly smaller volume at a given pressure than the 
experiment, while the GGA gives a larger volume.
  
\begin{figure}
\label{figPV}
\begin{center}
\includegraphics[height=80mm,width=110mm,angle=0]{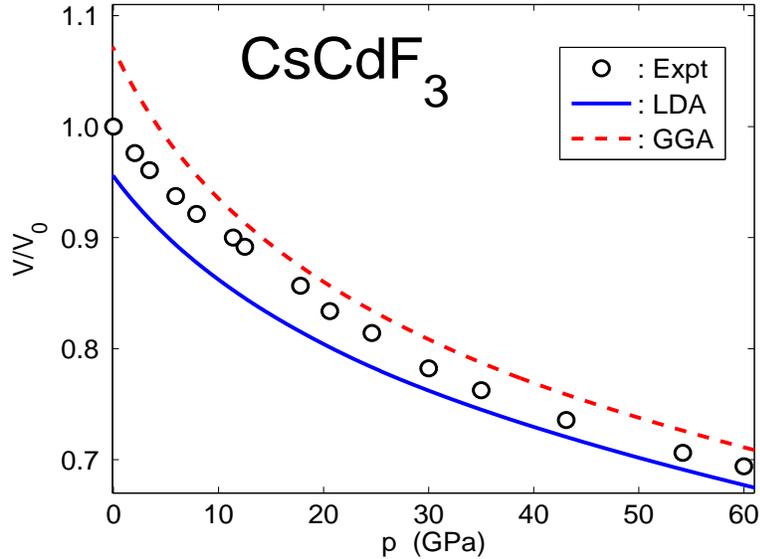}\\
\caption{ (Color online) Measured (circles) and calculated (full line: LDA, broken line: GGA) $pV$-relation
of CsCdF$_3$.}
\end{center}
\end{figure}

The lattice constant and bulk modulus measured in the present work as well as values
calculated within the LDA and GGA 
are given in Table I.
Results from earlier experimental works are also quoted for comparison.
The bulk modulus obtained in our experiments $B_0$ = 79(3) GPa with $B_0'$ =
3.8,  compares well with 
values calculated using the LDA. 
The lattice constant
obtained within the LDA is 1.6 \% lower than the experimental value,
while the corresponding bulk modulus is
4.7\% lower than the experimental value, which is the usual level of accuracy 
of LDA. When comparing the results
obtained within GGA, the lattice constant is 2.2 \% higher than the experimental value 
and the corresponding bulk modulus is 32\% lower than 
the experimental value.
Since the present theory does not describe thermal effects, it is 
more relevant to compare to the low temperature
experimental lattice constant, which is 0.4 \% lower than the room temperature (RT) value (Table I). Further, correcting
the low temperature experimental lattice constant for the zero-point
motion of the ions, which is not considered in the theory, reduces the best value to compare to theory even more. Hence, the 
LDA lattice constant is in fact closer to experiment   than the GGA value. 
 The excellent agreement between the 
 LDA calculated bulk modulus and the experimental 
 value is, however, fortuitous, since the bulk modulus depends strongly on volume. 
Due to the overestimated equilibrium volume within GGA (and underestimated 
with LDA), an error is introduced in 
the calculated bulk modulus. Therefore, we
recalculated the bulk modulus also at the experimental volume, using the simple scaling relation:\cite{Oxides}
\[
B(V)=B(V_0)\left(\frac{V_0}{V}\right)^{B'}.
\]
 The corrected values are also quoted in  Table I. 
We find that this diminishes 
the discrepancies between the LDA and GGA results, as expected. In addition, the LDA
bulk modulus now becomes {\it smaller}  than the GGA one for CsCdF$_3$, and both functionals are
seen to actually underestimate the bulk modulus, by approximately 25 \% (LDA) and 7 \% (GGA).

The elastic constants of CsCdF$_3$ calculated within LDA and GGA are 
listed in Table  2 where they are also
compared to experimental results as well as earlier calculations. The LDA overestimates the 
$C_{11}$ value by 28\% and the corresponding GGA value is 1.8\% lower than the experimental value. The 
$C_{12}$ value within LDA is 5.4\% lower than the experiment, and the corresponding GGA value is 33.3\% higher than the experiment. The
$C_{44}$ elastic constants are higher by  9\%  both within LDA and GGA compared to 
experiment.\cite{Rousseau}
The $C_{11}$ elastic constant obtained within GGA is much closer to 
the experimental value than the LDA value, while for $C_{12}$ the situation is reversed, and
for $C_{11}$ the two approaches give approximately the same value, 
and in reasonable agreement with the experimental value. 
The elastic constants 
depend sensitively on the volume
as illustrated in Figure 3 for the case of LDA. 
Hence the above values carry an appreciable uncertainty inherited from the volume inaccuracy of the LDA or GGA approach. The pressure derivatives
of the elastic constants  are more stable quantities. They are compared to experimental values\cite{Fischer}  in Table III with 
a reasonable agreement.
 The $C_{11}$ has the strongest pressure dependence in accordance with the experiment, while the theoretical $C_{44}$ value is almost constant with pressure,
while experiment find it increasing with pressure. 
A point of caution is the fact that the present theory pertains to $T=0$ K,
while experiments are performed at room temperature. Finite 
temperature generally tends to reduce the elastic constants because of thermal expansion. 
Using the calculated elastic constants we calculated 
the anisotropy factor $A = 2C_{44}/(C_{11} - C_{12})$.
We find an $A =   0.49$ for LDA and $A=0.70$ for GGA. 
The experimental\cite{Rousseau} value calculated from the elastic constants is 0.74 which is slightly higher than the GGA value.

\begin{figure}
\label{figC}
\begin{center}
\includegraphics[width=150mm,angle=0,clip]{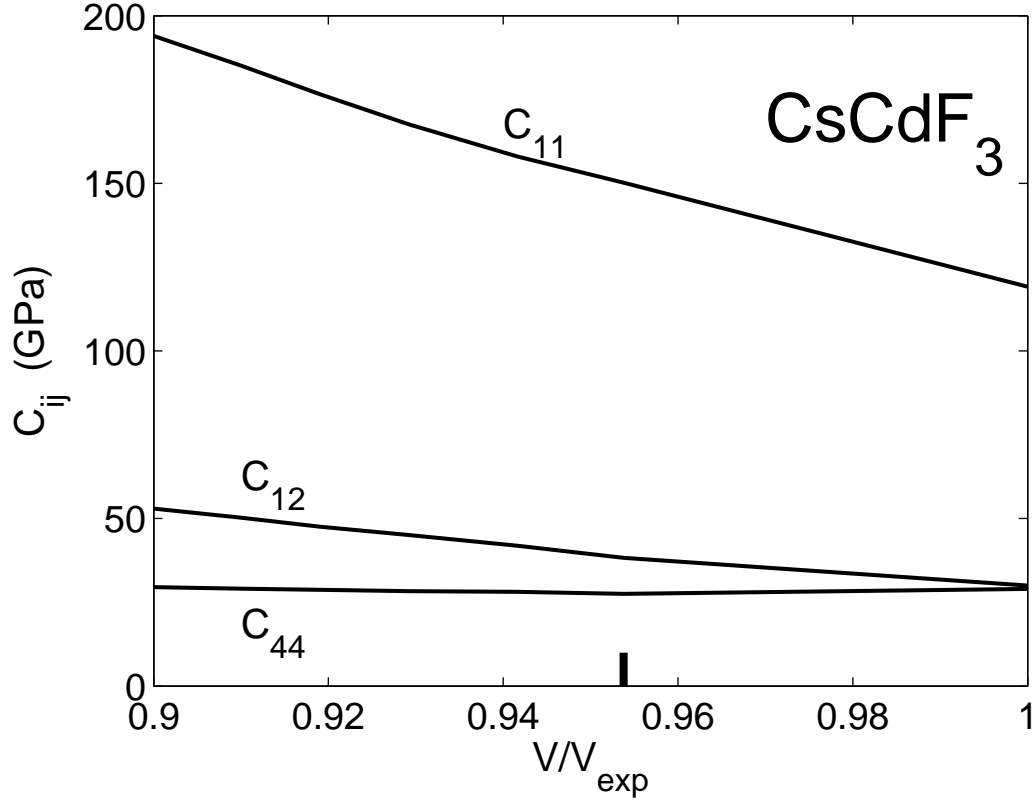}\\
\caption{Volume variation of the elastic constants of CsCdF$_3$ as calculated in LDA. The volume is given relatively
to the experimental volume, while             the theoretical LDA equilibrium volume is marked with a vertical bar on the first axis.
}
\end{center}
\end{figure}

\section{Electronic structure}

The calculated electron band structure of CsCdF$_3$ is shown in Figure 4, 
while the
 density of states is displayed in Figure 5. 
The valence bands consist of the fluorine 
p bands, which dominate the first ~2 eV below the valence band maximum. Below these bands appear the 
Cs 5p bands and the Cd 4d bands, which hybridize 
heavily in the region from 6 to 4 eV below the top of the valence band.
The Cd 6s states dominate the bottom of the conduction band.
The insulating gap is calculated to be 3.16 eV in LDA (3.67 eV in GGA).
With compression, the gap increases
almost linearly, at the rate
\[
V\frac{dE_g}{dV}=-2.1 \mbox{ eV}.\]
The conduction band minimum occurs at the $\Gamma$ point, while
the valence band maximum occurs at the R-point $(1/2,1/2,1/2)\frac{2\pi}{a}$, however almost
no dispersion is found for the uppermost valence band along the line connecting the 
R-point with the M-point $(1/2,1/2,0)\frac{2\pi}{a}$.
\begin{figure}
\label{figband}
\begin{center}
\includegraphics[width=100mm,angle=0,clip]{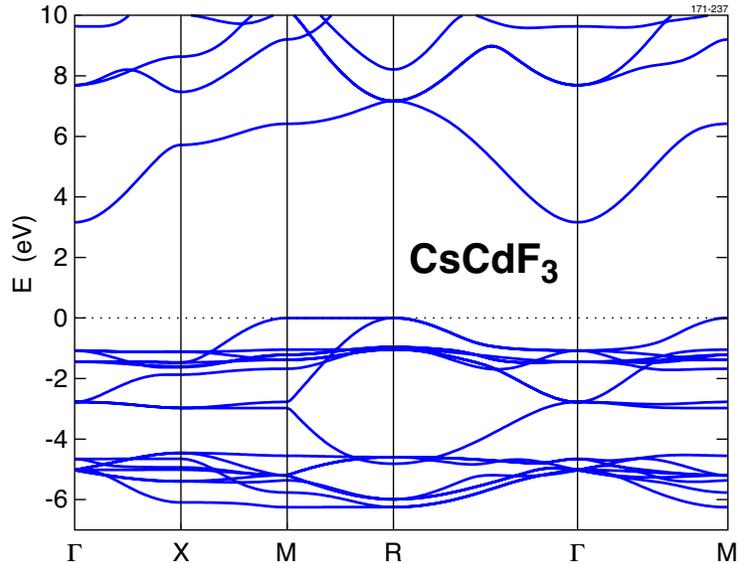}\\
\caption{Band structure of CsCdF$_3$ (using LDA, at the experimental lattice constant, $a=4.4669$ \AA).
The zero of energy is set at the position of the valence band maximum.
}
\end{center}
\end{figure}

\begin{figure}
\label{figdos}
\begin{center}
\includegraphics[width=100mm,angle=0,clip]{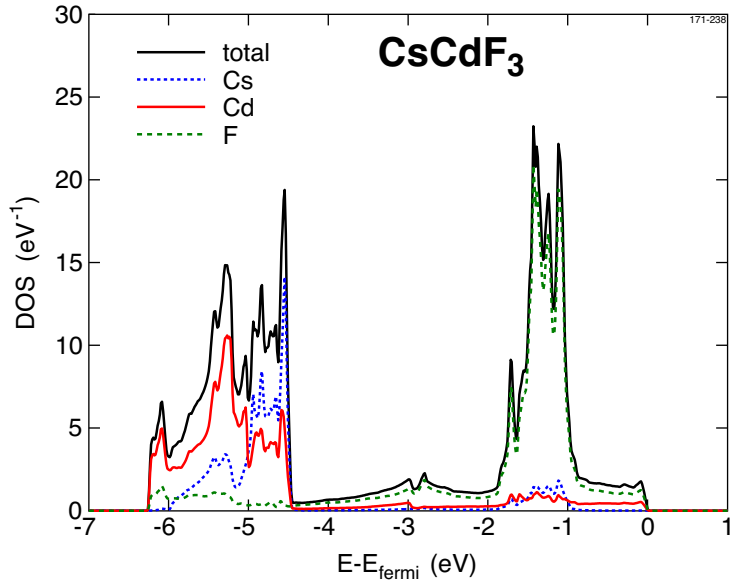}\\
\caption{(Color online) Valence band density of states of CsCdF$_3$ (using LDA, at the experimental lattice constant).
The zero of energy is set at the position of the valence band maximum.
The partial projections onto the spheres of Cs, Cd and F are shown with dotted (blue), dashed (red) and
dash-dotted (green) lines, respectively, while the full line gives the total density of states. Units
are electrons per eV and per  formula unit.
}
\end{center}
\end{figure}

\section{Conclusions}

In the present work, a combined theoretical and experimental analysis of the structural stability and equation of 
state of the fluoroperovskite CsCdF$_3$ has been carried out up to a pressure of 60\,GPa. We find that the compound remains 
in the cubic structure in the entire pressure range studied. 
The calculated equilibrium lattice constant, bulk modulus 
and elastic constants agree well with available experimental data. 
Our electronic structure calculations 
show that CsCdF$_3$ is an insulator with an indirect gap which increases with pressure.

\acknowledgments
G. V,  V. K acknowledge 
VR and SSF for financial support and SNIC for providing computer time. 
Part of the calculations were carried out at the Center for Scientific Computing in
Aarhus (CSCAA) supported by the Danish Center for Scientific Computing. 
The authors gratefully 
acknowledge the use of the HPCAT facility supported by DOE-BES, DOE-NNSA, NSF, and 
the W. M. Keck Foundation. HPCAT is a collaboration among the
Carnegie Institution, Lawrence Livermore National Laboratory, the University 
of Nevada Las Vegas, and the Carnegie/DOE Alliance Center. We thank Dr. Maddury Somarazulu and 
other HPCAT staff for technical assistance. This research was supported from the 
U.S. Department of Energy Cooperative Agreement No. DE-FC52-06NA26274 with the University of Nevada, Las Vegas.  
\clearpage

\begin{table}[tb]
 \caption{
Lattice constants (in \AA), bulk modulus $B_0$ (in GPa)
 and its pressure derivative $B_0'$,
 of CsCdF$_3$ as obtained by experiment and theory. The bulk moduli have been calculated both at the experimental and theoretical volumes
($B_0(V^{\rm{exp}}_{0})$ and $B_0(V^{\rm{th}}_{0})$, respectively)}
\begin{ruledtabular}
\begin{tabular}{cccccc}
& Lattice constant &   $B_0(V^{\rm{th}}_{0})$  & $B_0(V^{\rm{exp}}_{0})$     & $B_0'                 $  \\
\hline
	
       GGA$^a$ & 4.567     & 53.3    &  73.6      &  4.9  \\
       LDA$^a$ & 4.397     & 75.6   &   59.7      &  4.9 \\
       Expt.(RT)   & 4.4669(7)$^a$, 4.465$^b$, 4.4662$^c$, &       & 79(3)$^a$   &  3.8$^a$,5.8$^d$  \\
       Expt.($T=0$ K)   & 4.452$^b$ &       &   &   \\
    
\end{tabular}
\end{ruledtabular}
$^a$Present work; 
$^b$Ref. \onlinecite{Rousseau};
$^c$Ref. \onlinecite{Damodar}; 
$^d$Ref. \onlinecite{Fischer}. 

\end{table}

\begin{table}[tb]
\caption{
Calculated elastic constants and  shear modulus (G), all expressed in GPa, 
of CsCdF$_3$
at the theoretical equilibrium volume.}
\begin{ruledtabular}
\begin{tabular}{cccccc}
              &$C_{11}$  &$C_{12}$  &$C_{44}$ &  $G$ &  \\ \hline

GGA           & 105.8     & 27.0      &  27.7             & 32.4    & Present    \\         
LDA           & 150.2    & 38.3     &  27.5              & 38.9     & Present   \\
Expt.         & 107.8$\pm$0.2  & 40.5$\pm$0.5  &  25.0$\pm$0.2  & 28.5$\pm$0.5 &  Ref.  \onlinecite{Rousseau} \\
\end{tabular}
\end{ruledtabular}
\end{table}

\begin{table}[tb]
\caption{
Calculated LDA pressure derivatives of the elastic constants 
of CsCdF$_3$ at the experimental equilibrium volume.}
\begin{ruledtabular}
\begin{tabular}{ccccc}
              &$\frac{dC_{11}}{dp}$  &$\frac{dC_{12}}{dp}$  &$\frac{dC_{44}}{dp}$ &   \\ \hline

LDA           &  8.8     &  2.9     &   0.4                     & Present   \\
Expt.         &  11.1$\pm$0.1  &  3.2$\pm$0.1  &  2.2$\pm$0.4  &  Ref.  \onlinecite{Fischer} \\
\end{tabular}
\end{ruledtabular}
\end{table}

\end{document}